# Proposal and design of 81.25 MHz heavy ion drift tube linacs for BISOL

Meiyun Han [1], Tianhao Wei [1], Ying Xia [1], Austin Morris[1] ,Yuanrong Lu [1],*, Zhi Wang[1],*, Zhaohua Peng [2]

1 State Key Laboratory of Nuclear Physics and Technology, Peking University, No. 201, Chengfu Road, Haidian
District, Beijing, 100871 China

2 Department of Nuclear Physics, China Institute of Atomic Energy, Xin Town, Fangshan District, Beijing,
102413 China

* Correspondence: yrlu@pku.edu.cn (Y.R.Lu), wangzhi@pku.edu.cn (Z.Wang).

## Abstract

Based on the design requirements proposed by the Beijing On-Line Isotope Separation project (BISOL), four $Sn^{22+}$-based, 81.25 MHz continuous wave (CW) drift tube linac (DTL) cavities have been designed. These DTLs are capable of accelerating $Sn^{22+}$ of 0.1 pmA from 0.5 MeV/u to 1.8 MeV/u over a length of 7 m, with an output longitudinal normalized RMS emittance of 0.35π·mm·mrad, and transmission efficiency higher than 95%. The dynamics design adopted the KONUS (Kombinierte Null Grad Struktur Combined 0° Structure) scheme. Comprehensive error study implies that these DTLs can accommodate a wide range of non-ideal beams and cavity alignment errors while maintaining high transmission efficiency. The electromagnetic design employed a Cross-bar H-mode (CH) structure for superior water-cooling characteristics, and a detailed tuning analysis was conducted to derive an optimal tuning scheme. The results of the multiple-physics analysis indicate that the frequency shift of each cavity is within an acceptable range. Comparing the dynamics requirements with the RF design results, similar particle output phase distribution, equivalent energy gain and consistent emittance growth are observed. Detailed designs will be presented in this manuscript.
Key words: BISOL, CW, CH drift tube linacs

## 1 Introduction

The BISOL [1] is an emergent scientific facility jointly applied by Peking University and the China Institute of Atomic Energy (CIAE), which will use a combination of Isotope Separation On-Line (ISOL) and the Projectile Fragmentation (PF) [2] methods to generate high-intensity, neutron-rich, radioactive particle beams for advanced nuclear physics research. As shown in Fig.1, primary radioactive particles accelerated by the post-acceleration system will be generated in two ways, the first method being front-end deuteron accelerator and reactor system generation, and the second method being direct generation from the Electron Cyclotron Resonance (ECR) ion source. The post-acceleration system are proposed to accelerate the particles to 150 MeV/u, and collided with a production target to form secondary, extremely neutron-rich ion beams.

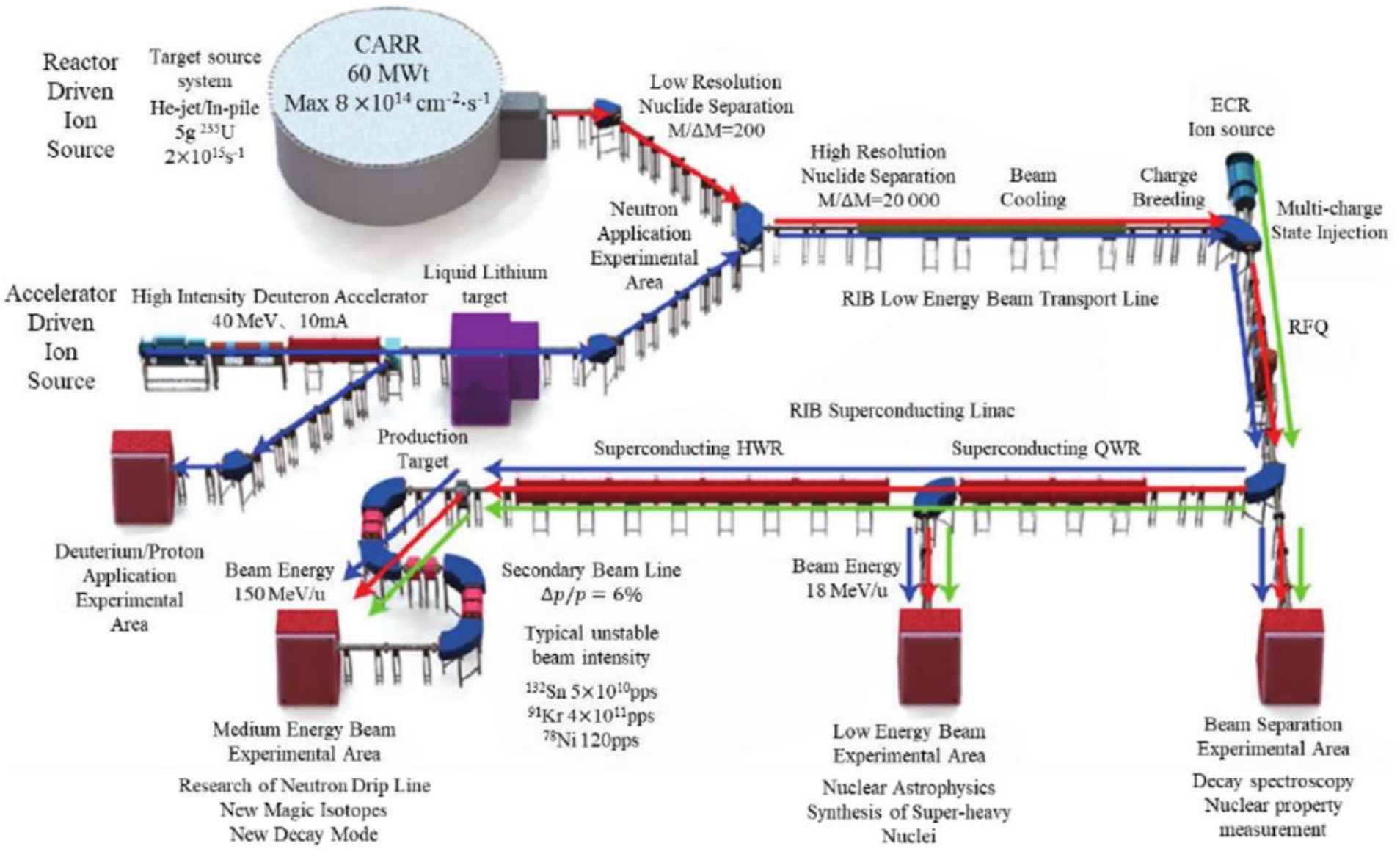

Figure 1. Layout diagram of BISOL [1]

The post-accelerator system is primarily composed of low-energy beam transport (LEBT), radio frequency quadrupole (RFQ), medium-energy beam transport (MEBT), superconducting linear accelerator and high-energy beam transport (HEBT) systems. The initial strategy was to inject the beam directly into the superconducting linacs at the exit of the MEBT. Since increasing the initial beam energy injected into the superconducting linacs can substantially reduce the number of superconducting cavities, the design of the DTLs positioned before the superconducting linacs was proposed instead. This design requires the final output energy of the DTL to be no less than 1.8MeV/u. As a CW device, the DTLs in the BISOL post-accelerator system must exhibit a higher shunt impedance, lower energy consumption, and posess an optimized water-cooling structure.

Despite promising theoretical results, the international construction and operation of room temperature DTL tanks are relatively scarce. CH structures operating at such low frequencies are even less prevalent. Nonetheless, several projects have provided valuable insights and significantly contributed to the design proposed in this paper, including but not limited to FAIR DTL [3] [4] [5], MYRRHA DTL [6] [7], KHIMA DTL [8], JPARC DTL [9], and SNS DTL [10] [11].

## 2 Beam dynamics design

The beam dynamics design was implemented utilizing the LORASR code [12], following the KONUS beam dynamics proposed by U.Ratzinger [13]. A KONUS period consists of transverse focusing by a quadrupole triplet (QT) or solenoid, longitudinal focusing by 2 to 7 drift tubes at negative synchronous phase $\varphi_s$, and main acceleration along a multi-gap structure at 0°.

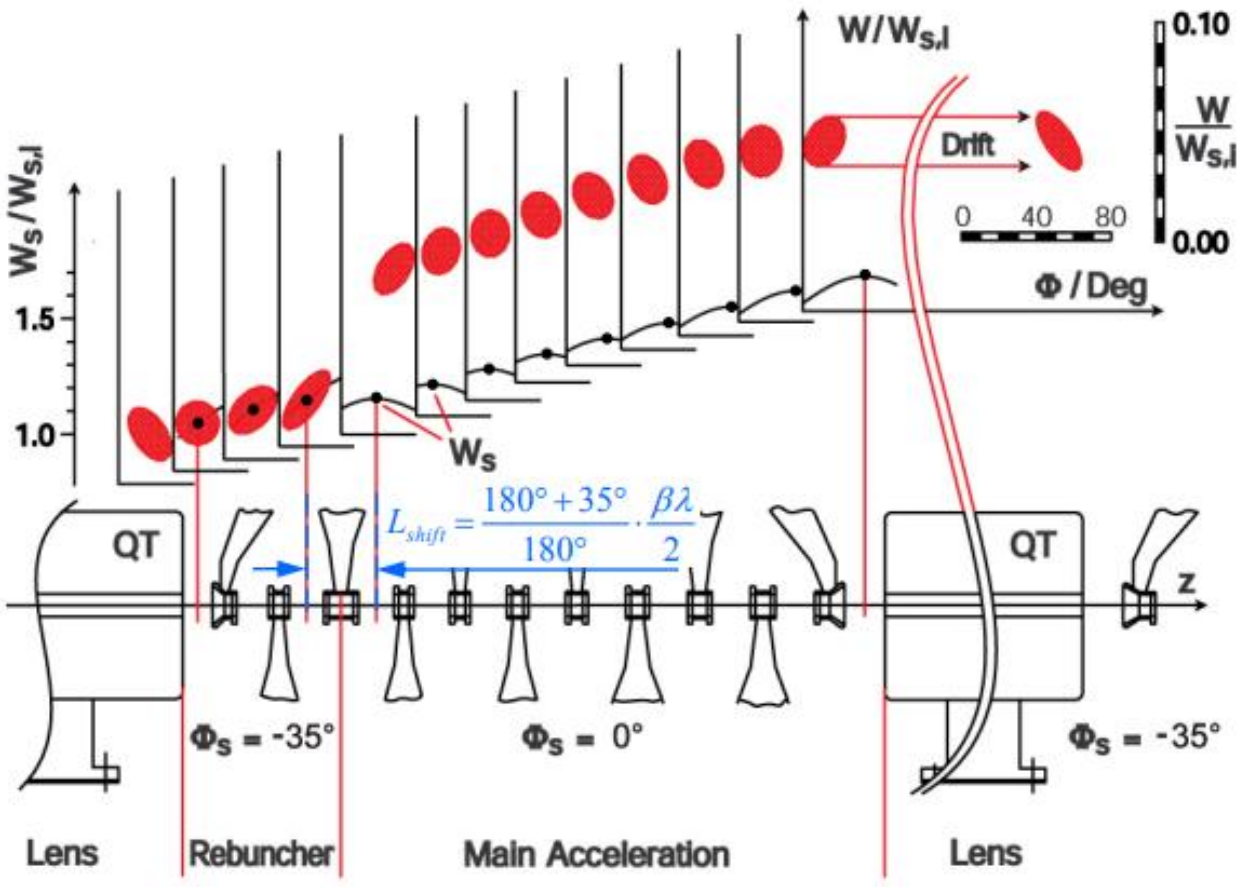

Figure 2. A KONUS period [14]

The output particle distribution of the MEBT serves as the input for the DTL, as illustrated in Fig.3. Similar to the completed RFQ [15], the DTL tanks also operate at 81.25 MHz, with $Sn^{22+}$ being the particle of choice corresponding to a 0.1 pmA peak beam current. The design requires a transmission efficiency exceeding 95%, an output energy of no less than 1.8 MeV/u, and minimal growth of transverse normalized RMS emittance.

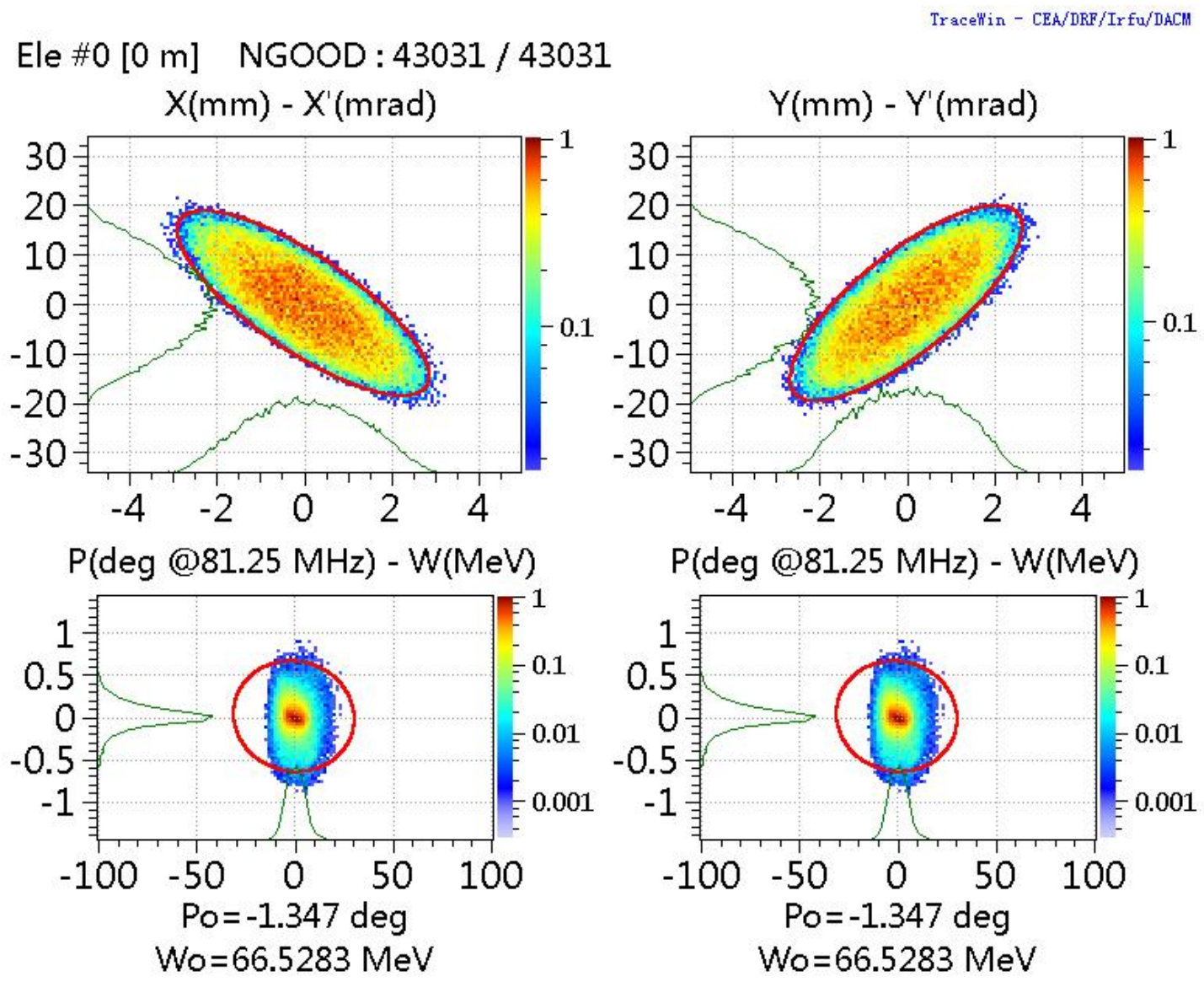

Figure 3. Beam output distribution of MEBT

The design process necessitates numerous iterations of beam dynamics and RF calculations. In order to achieve a small envelope with high transmission efficiency, it is imperative to identify suitable values for the synchronous phase, gap voltage distribution, drift tube length, and gap length for each accelerating cell. Subsequently, an RF model is constructed and calculated using CST Micro Wave Studio [16], based on the result in LORASR. The structure parameters are next adjusted until the electric field distribution is flat and the integral voltage of the gaps is as close to the LORASR result as possible. Finally, particle tracking with RF field maps from CST is performed to authenticate the beam dynamics design. If the simulation results satisfy the beam dynamics requirements, then the iterative process is complete. If the beam distribution at the exit deviates significantly, however, then another iteration cycle is instigated. After several iterations, the main parameters of the dynamics results are shown in Table 1.

Table 1 Main parameters of dynamics results

| Parameters | value |
|---|---|
| Frequency [MHz] | 81.25 |
| Reference particle | $Sn^{22+}$ |
| Peak beam current[pmA] | 0.1 |
| Duty factor [%] | 100 |
| Input energy [MeV/u] | 0.504 |
| Output energy [MeV/u] | 1.81 |
| Input transverse normalized RMS emittance[$\pi$·mm·mrad] | 0.21 |
| Output transverse normalized RMS emittance[$\pi$·mm·mrad] | 0.35 |
| Number of gaps | 43 |
| Number of 0° gaps | 32 |
| Negative phase/deg | -40/-35 |
| Number of QT | 4 |
| Total length[mm] | 700.00 |

The design comprises four KONUS sections, including four cavities, 43 gaps, and four QTs, positioned between cavities. The number of gaps in each cavity is 7,11, 13 and 12, respectively. The beam energy is shown in Fig.4 and the envelope distribution is shown in Fig.5. The inner radius of the tubes is larger than 13.0 mm, the maximum envelope radius within the tubes is less than 8.0 mm, and the aperture margin is almost 40%, avoiding the space-charge effect induced by halo particles.

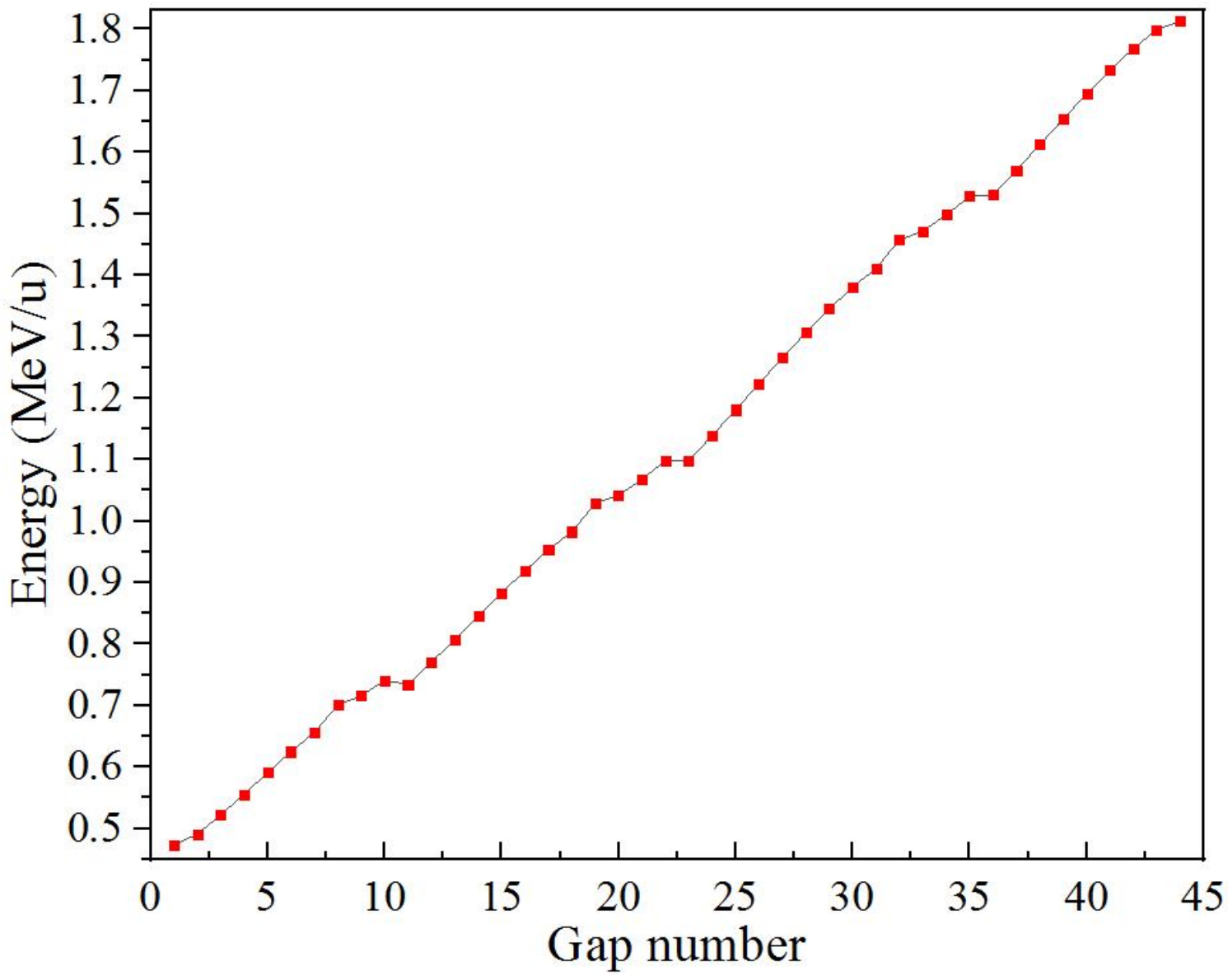

Figure 4. Energy in LORASR

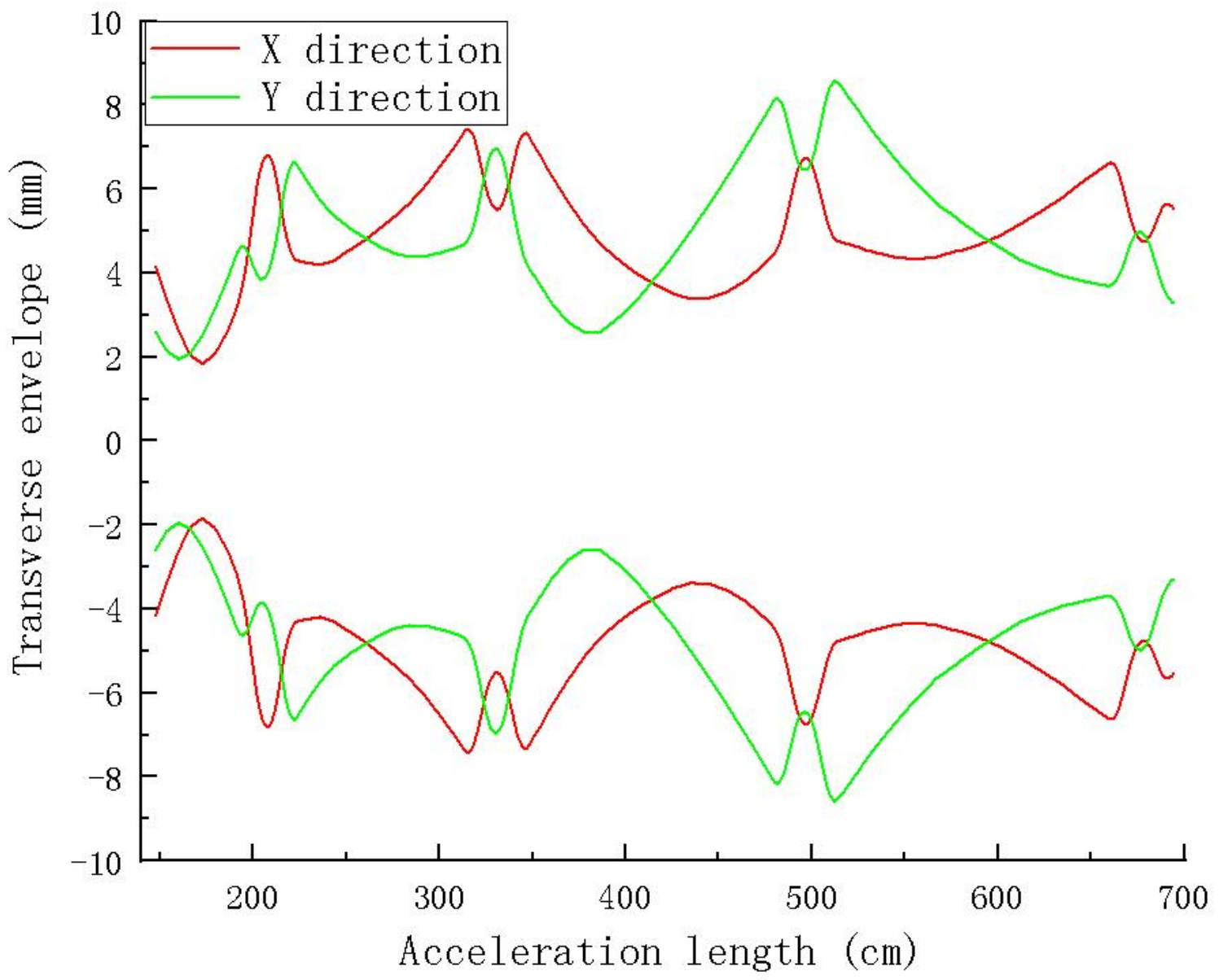

Figure 5. Transversal envelope in LORASR

The $\varphi_s$ within the rebunching section, -40° and -35°, provide a large longitudinal acceptance, required by the tanks, and a strong longitudinal bunching at low energy. The final energy reaches 1.81 MeV/u, as the longitudinal normalized RMS emittance grows to 0.35 $\pi$ ·mm·mrad, and the transmission efficiency of the dynamics simulation reaches 100%.

## 3 Electromagnetic design and multiple-physics analysis

As a commonly employed DTL with high shunt impedance [17], the CH type has superior water-cooling characteristics compared to the IH type, and thus advantageous for CW operation.

Based on the dynamics design, 3D cavity models were built using CST. To ensure precision, mesh benchmarking was conducted prior to starting the electromagnetic field calculation.

Since the electric field is concentrated around the drift tubes, the mesh is encrypted there until the frequency converges.

## 3.1 Electromagnetic design result

### 3.1.1 Design of DTL1

DTL1 was initially designed with a coupled structure (CCL) [18], as shown in Fig.6, where a triplet is placed inside the coupling section to ensure transverse beam focusing. As described in the dynamics design section, there are 18 gaps in DTL1, 7 in the first CH resonator and 11 in the second one. The length of the first resonator is 0.75 meters and the second resonator is 1.01 meter.

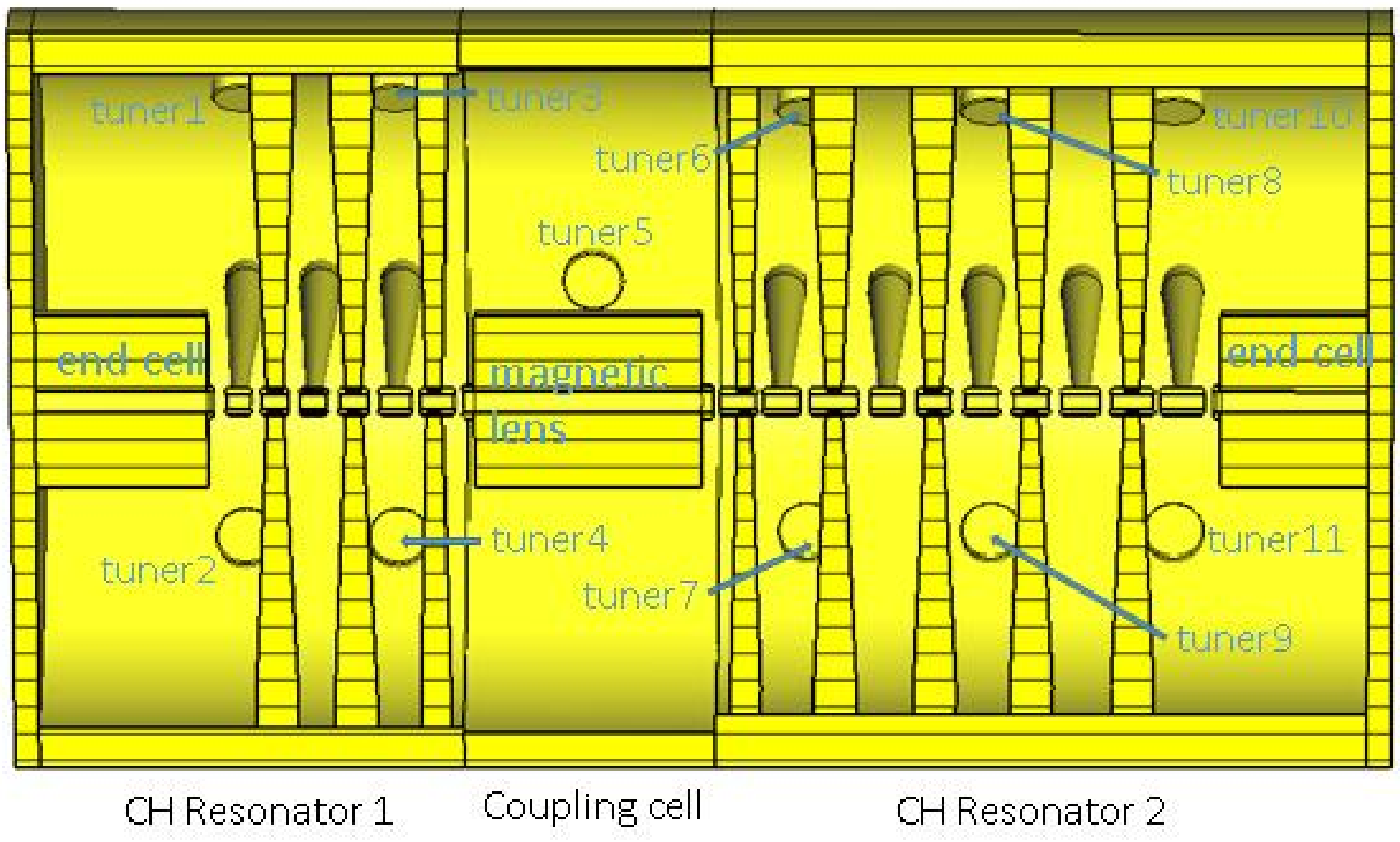

Figure 6. Schematic of initail design for DTL1

During the simulation, it was found that the mode sparation was kept around 0.2 MHz when varying the dimensions such as the insertion depth of the tuner and the end cells, such a small mode separation will challenge the power source. Therefore, DTL1 was finally divided into two normal DTLs with similar structure, as shown in Fig.7.

The peak surface electric field $E_S$ is a crucial constraint of the cavity design, since excessive $E_S$ may result in electrical breakdown or sparking. Results of the simulation revealed that $E_S$ is located at the edges of the drift tube, and can be reduced and optimized by increasing the blend radius and gap length. For the optimized design, the $E_S$ of the first tank was 17.4 MV/m located at the outer edge of the third tube with 6.0 mm blend radius, and the $E_S$ of the second tank was 15.4MV/m located at the outer edge of the sixth tube with 7.0 mm blend radius. Based on the empirical formula $f = 1.643{E_K}^2 e^{\left(-\frac{8.5}{E_k}\right)}$ [19], where f is the cavity frequency in MHz and $E_k$ is the Kilpatrick peak surface field in MV/m, when $E_S$ is lower than $1.8E_K$[20], it is conducive to CW operation. The $E_S$ of the two tanks are $1.65E_K$ and $1.46E_K$ at 81.25 MHz, respectively, the design criteria is met. Normally, a lower designed beam energy results in a higher cavity shunt impedance [18]. However, the design of increasing the gap length to reduce the $E_S$ resulted in a decrease in the Transition Time Factor TTF of each gap, leading to the lowest shunt impedance for Tank1

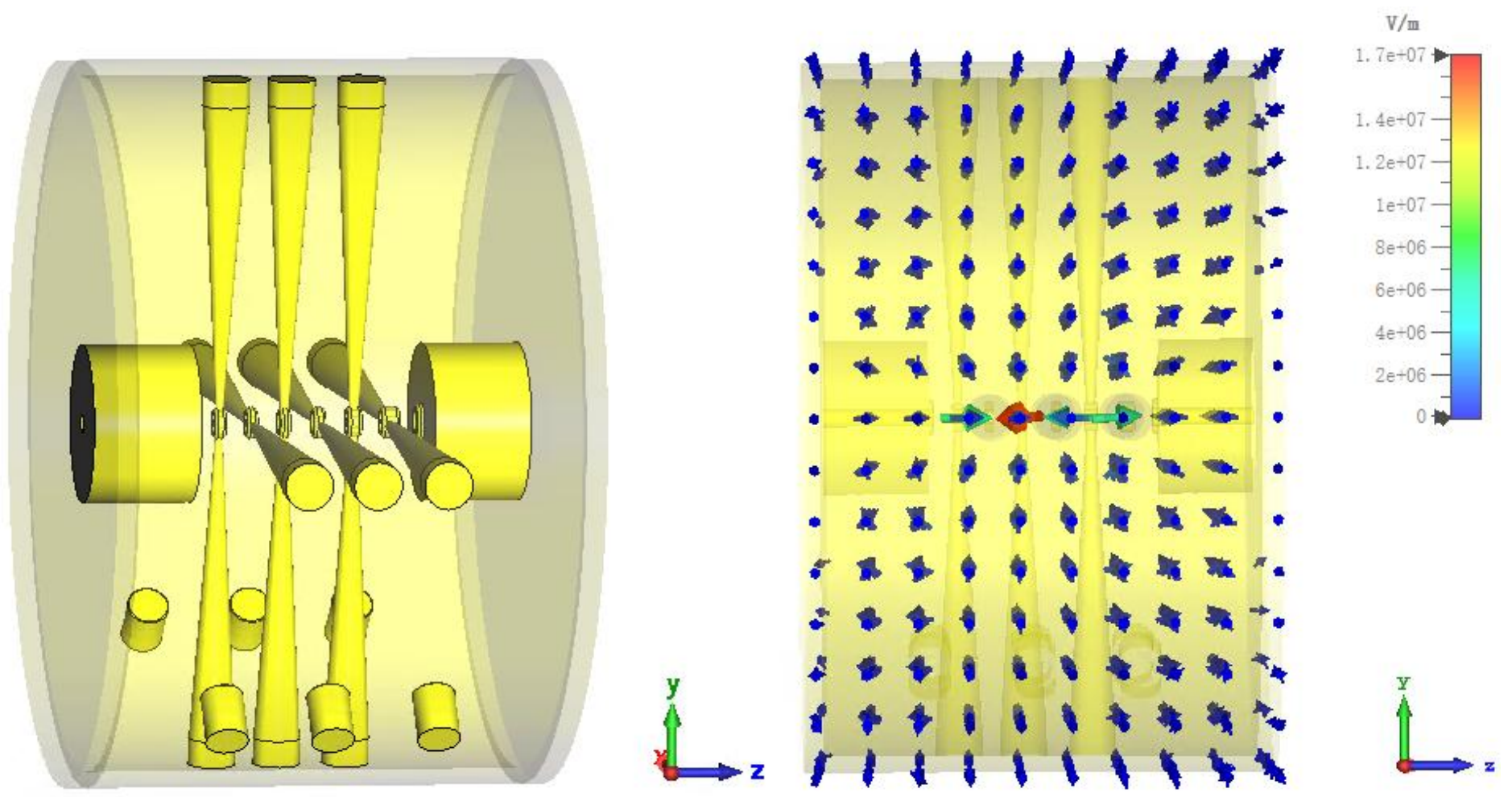

Figure 7 Structural schematic and electromagnetic distribution of DTL1-tank1

The final design incorporated 7 gaps into the first tank and 11 into the second. The Q-values for the two cavities were 20785 and 21300, respectively. All tuners have an initial insertion depth of 40.0 mm, as the insertion depth increases linearly from 0.0 mm to 70.0 mm, the tuning frequency range of the first tank with 6 tuners was 81.25 ± 0.04 MHz, and for the second tank with 8 tuners, the range was 81.25 ± 0.05 MHz. The frequency variation of the cavity was approximately linear with the insertion depth of the tuners. The primary parameters of the cavity design are presented in Table 3. The mode separations of both tanks exceed 15 MHz, indicating an advantage for the uncoupled structure.

Table 3 Main parameters of DTL1-tank1 and DTL1-tank2

| Parameters | Tank1 | Tank2 |
|---|---|---|
| Length/m | 0.91 | 1.14 |
| Cavity power/kW | 26.1 | 35.7 |
| $E_S$ [MV/m] | 17.4 | 15.4 |
| Q | 20785 | 21300 |
| Number of tuners | 6 | 8 |
| Radius of tuners/mm | 40.0 | 40.0 |
| Frequency range/MHz | 81.24-81.29 | 81.19-81.30 |
| Mode separation/MHz | 24.5 | 17.0 |
| Input energy[MeV/u] | 0.504 | 0.706 |
| Output energy[MeV/u] | 0.706 | 1.028 |
| Shunt impedance[MΩ/m] | 66.1 | 91.7 |

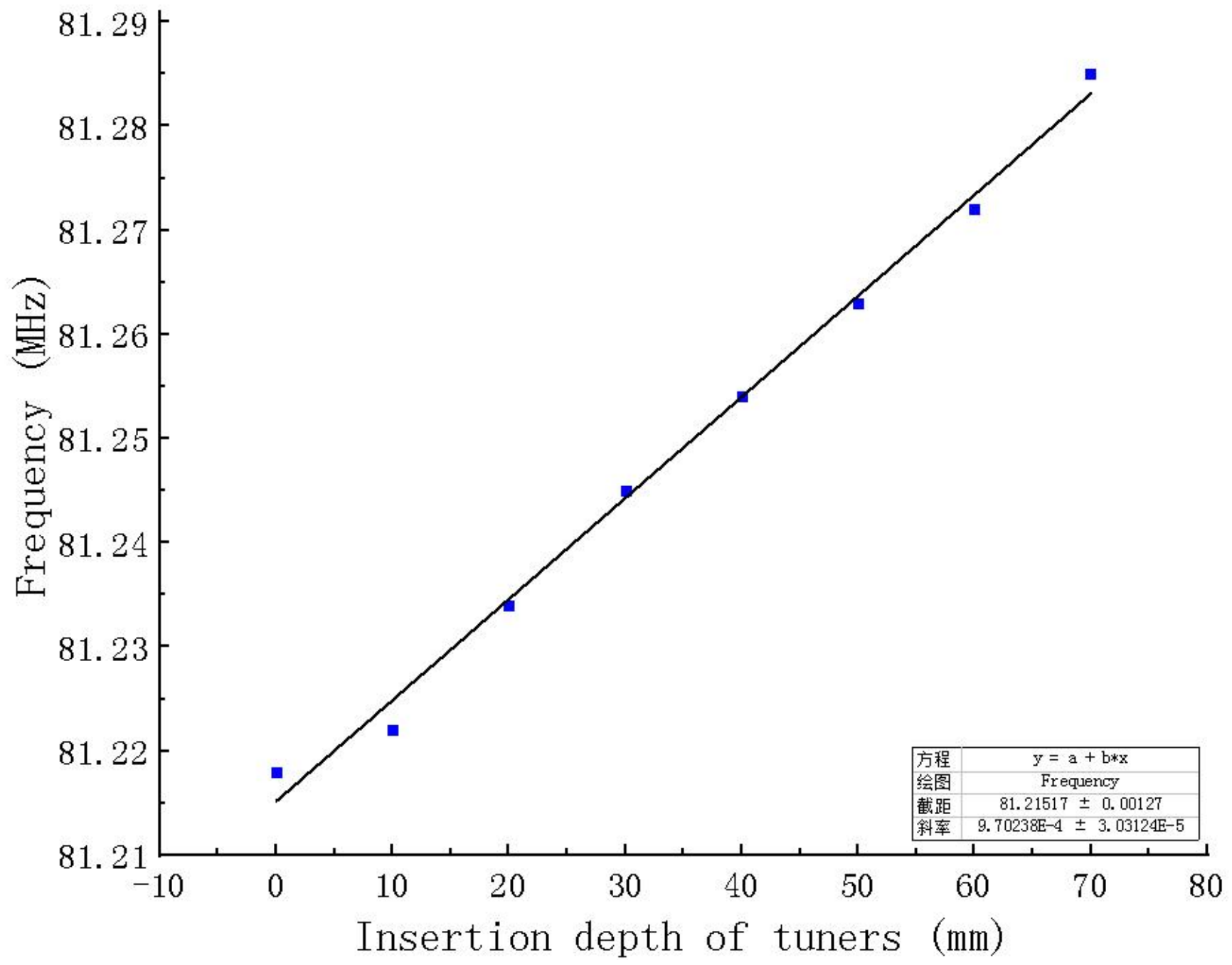

Figure 8 Frequency of DTL1-tank1 with different tuners insertion depth.

### 3.1.2 Design of DTL2 and DTL3

DTL2 and DTL3 employed the same uncoupled structure as previously described. Following an identical optimization , the $E_S$ of DTL2 and DTL3 were found to be less than 1.8$E_K$, and the cavity had a sufficient tuning ability. Parameters calculated by CST are shown in Table 4.

Table 4 Main parameters of DTL2 and DTL3

| Parameters | DTL2 | DTL3 |
|---|---|---|
| Cavity power/kw | 39.5 | 37.3 |
| Q | 21671 | 21042 |
| $E_S$ [MV/m] | 15.4 | 14.8 |
| Number of tuners | 10 | 10 |
| Radius of tuners/mm | 40.0 | 40.0 |
| Initial depth of tuners/mm | 40.0 | 40.0 |
| Insertion depth range of tuners/mm | 0.0-70.0 | 0.0-70.0 |
| Frequency range/MHz | 81.17-81.32 | 81.19~81.32 |
| Mode separation/MHz | 8.95 | 8.23 |
| Total length/m | 1.79 | 1.79 |
| Input energy[MeV/u] | 1.028 | 1.44 |
| Output energy[MeV/u] | 1.44 | 1.823 |
| Shunt impedance[MΩ/m] | 86.8 | 78.6 |

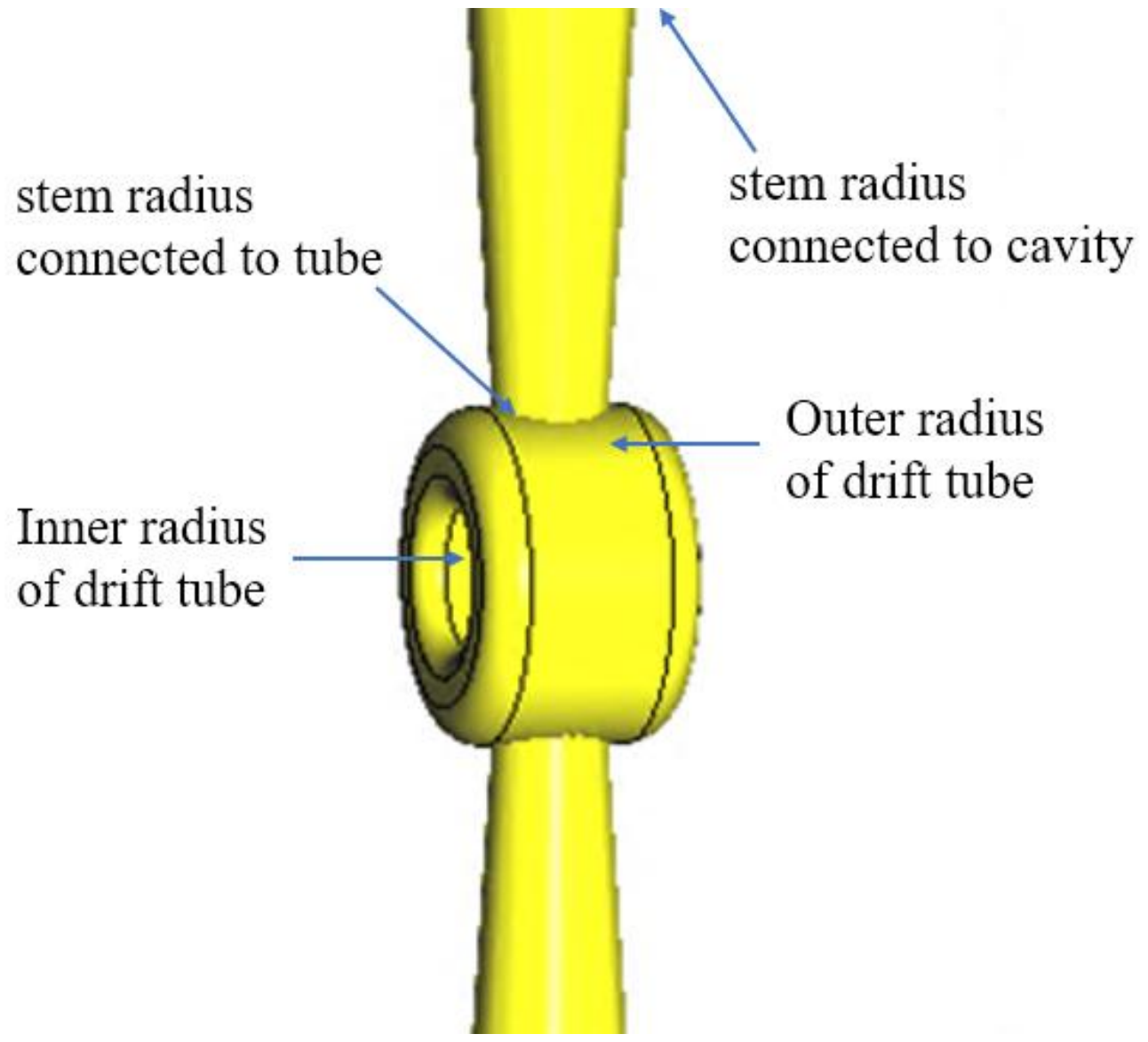

Figure 9 Design of drift tube and stem.

For DTL3, the impact of various drift tube and stem radii on the cavity parameters was studied. The structure is shown in Fig.9, for a cavity radius adjusted to tune the frequency to 81.25 MHz. As displayed in Fig.10, an increase in the inner radius of the drift tubes moderately augmented shunt impedance and Q, without producing significant alterations to $E_S$. Conversely, increasing the outer radius of the drift tubes markedly reduced the cavity size, $E_S$, and unit processing costs. However, since a significantly lower shunt impedance and Q is also anticipated in this case, it is necessary that the outer radii of the drift tubes be optimized with this in mind.

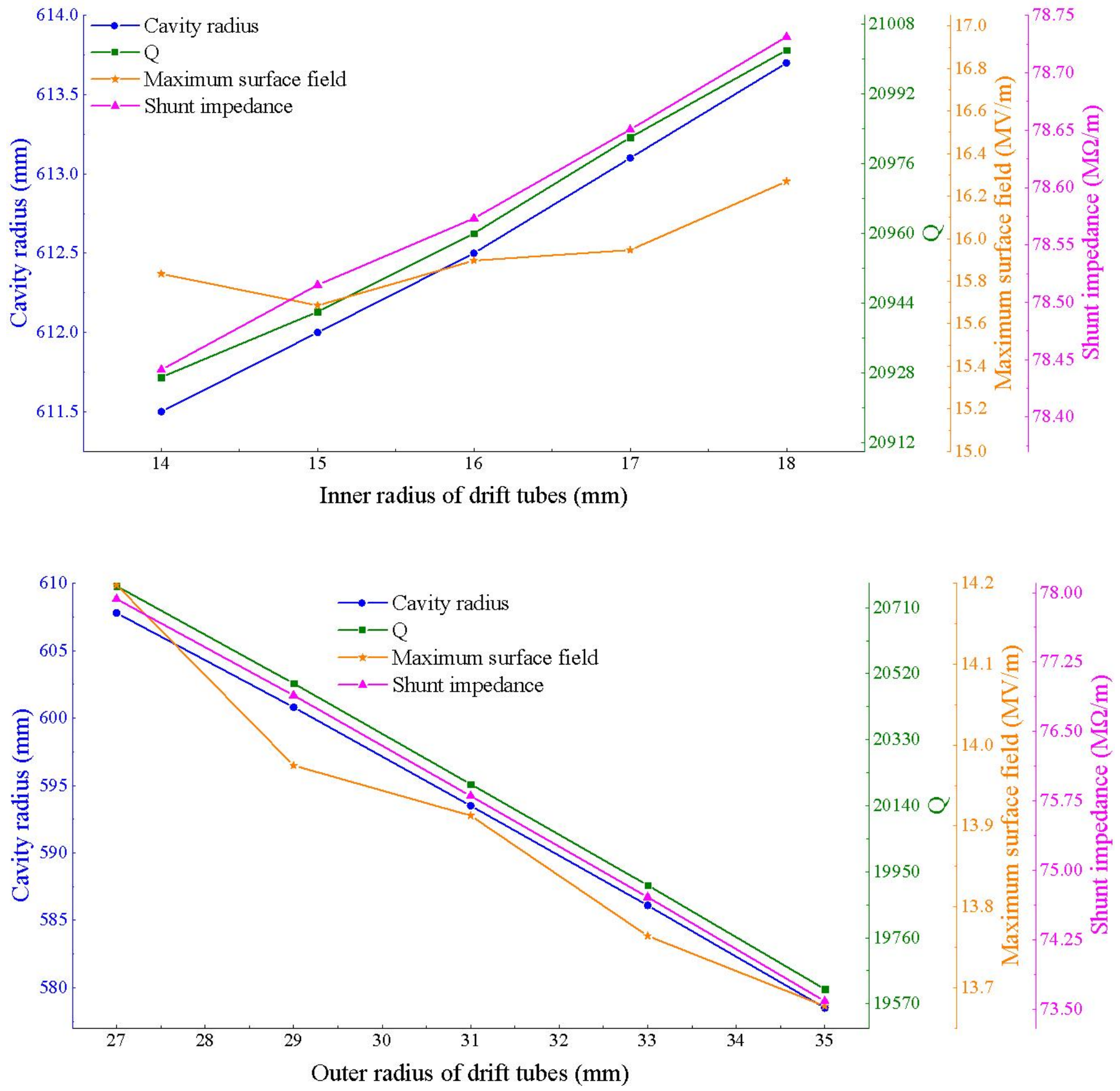

Figure 10 The relationship between the shunt impedance (red), peak surface electric field $E_S$ (yellow), Q, and tube radii. Here, the cavity radius was adjusted to tune the frequency of the DTL cavity to 81.25 MHz.

As illustrated in Fig.11, an increase in the value of the stem radius connected to the tube, resulted in a reduction in both $E_S$ and the cavity radius, while slightly increasing the shunt impedance and Q. This suggests that it is advantageous to increase the stem radius connected to the tube proportionally. Although the shunt impedance and Q are increased with the enlargement of the stem radius connected to the cavity, the slight increase in the cavity radius poses a minor variation to its machining.

The optimization of the drift tubes and stems incorporated this variation and considered the blending space for the edges of the drift tubes, and whether the stems are strong enough to support the drift tubes, etc. Therefore, the parameters of the drift tubes and stems vary across each DTL, but the structural features remain similar.

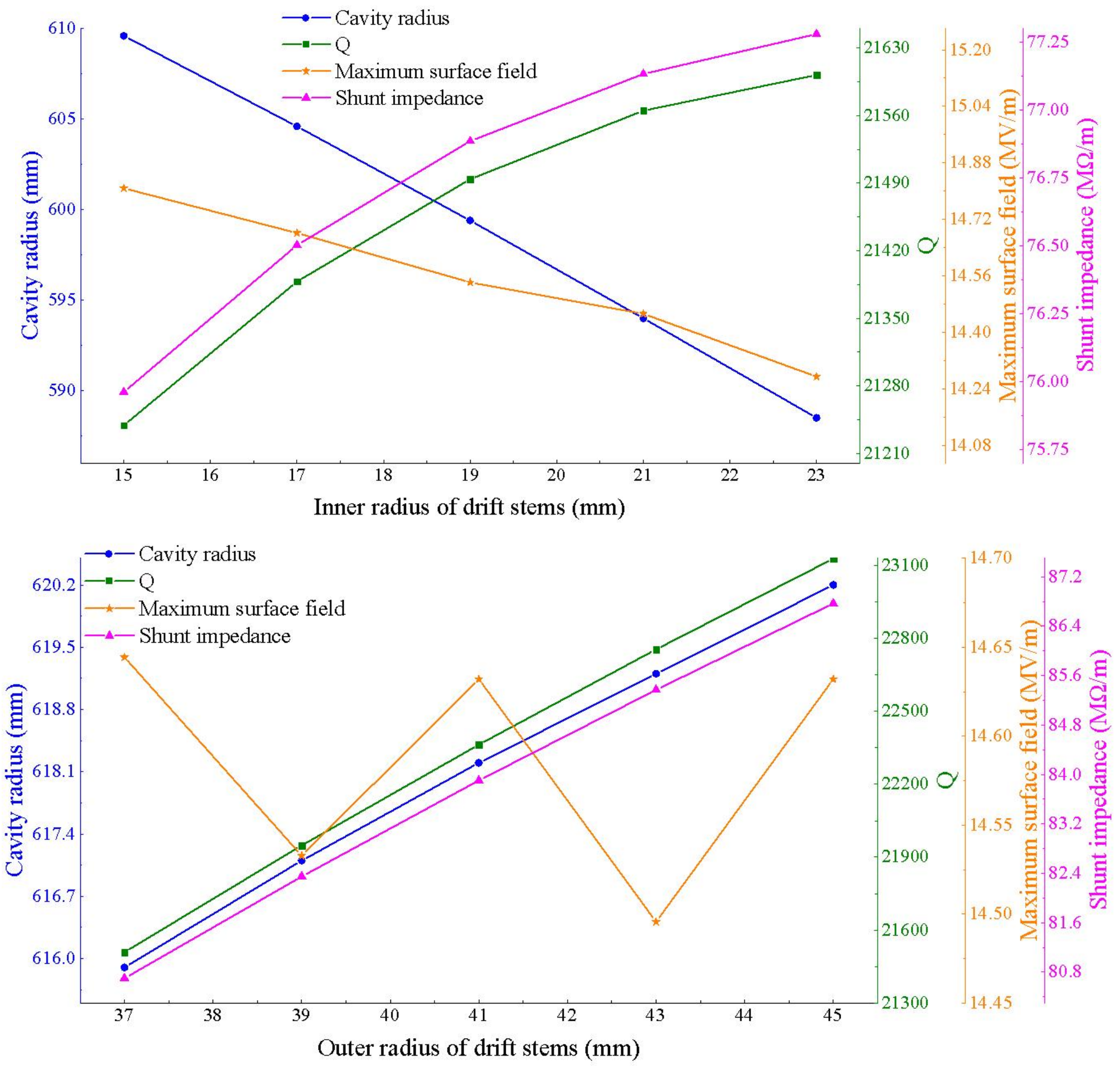

Figure 11 The relationship between the shunt impedance (red), peak surface electric field $E_S$ (yellow), Q, and stem radii. Here, the cavity radius was adjusted to tune the frequency of the DTL cavity to 81.25 MHz.

## 3.2 Multiple-physics analysis

As cavity consumption will cause structural deformation and a frequency shift, efficient distribution of the waterpipe cooling distribution can be used to lower the deformation and frequency shift to an acceptable level. The subsequent multiple-physics simulation was performed using CST. The waterpipe distribution was consistent for each DTL, as shown in Fig.12. Each stem-couple and corresponding drift tube constituted a water channel, simplifying the design of the cooling system. One water channel was set on the outer wall of the magnet and 16 water channels were set on the outer wall of the cavity.

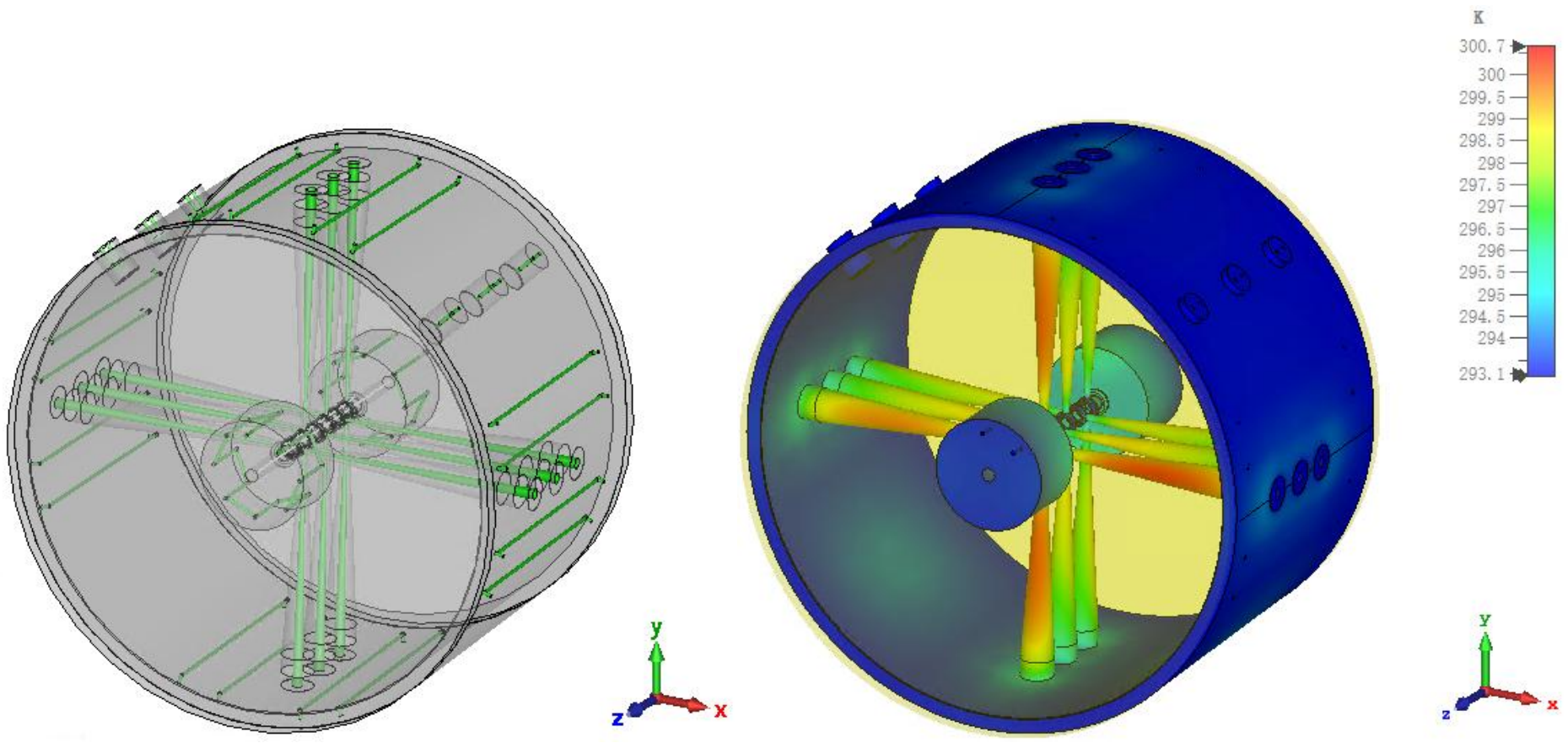

Figure 12. Waterpipe cooling distribution of DTL1-tank1 and temperature distribution

In the simulation, both the water and ambient temperature were set to 20°C. The heat transfer coefficient h can be obtained by the Dittus-Boelter correlation [21]:

$$h=\frac{N_u k}{D},$$

where k = 0.6 W/mK is the water heat conductivity coefficient and D is the diameter of the cooling channel. Nu is the Nusselt number, which is defined as:

$$N_u = 0.023 \cdot Re^{0.8} \cdot Pr^{0.4}$$

$$P_r = \frac{C_p \mu}{k}$$

$$R_e = \frac{vD\rho}{\mu},$$

where v is the water velocity, μ=0.001 Ns/m$^2$ is the viscosity of water, Cp=4.2 kJ/kg is the heat capacity of water, ρ=998 kg/m$^3$ is the density of water, and Re and Pr are the Reynolds and Prandtl numbers. The velocity of the water was assumed to be 2.5 m/s.

At the stem, the maximum temperature rise of tank1 was observed to be 7.4℃. The maximum deformation, with a size of 0.16 mm, occurred in the first tube, with a corresponding maximum thermal stress of 47.9 MPa, which is beneath the limit of oxygen-free copper (60 to 70 MPa) [22]. The frequency shift caused by the deformation was -5.8 kHz. The results for each cavity are summarized in Table 5, for which the maximum temperature rise and thermal stress of each cavitity is given. The maximum deformation of DTL1-tank1 and DTL1-tank2 occured on the drift tubes, due to their short length, while the maximum deformation of DTL2 and DTL3 occured on the stems. Nonetheless, the consequent frequency shift was within an acceptable range.

Table 5 Multi-physics results of cavities

| Parameters | DTL1-Tank1 | DTL1-Tank2 | DTL2 | DTL3 |
|---|---|---|---|---|
| Maximum temperature rise/℃ | 7.4 | 6.7 | 7.5 | 7.67 |
| Maximum displacement/mm | 0.160 | 0.107 | 0.075 | 0.071 |
| Maximum thermal stress/MPa | 47.9 | 25.7 | 21.1 | 27.8 |
| Frequency shift/kHz | -5.8 | -8.5 | -3.7 | -3.6 |

# 4 Error study

## 4.1 Comparison of dynamics and RF simulation results

For any accelerator design, variations between dynamics and RF simulations arise inevitably. The primary requirement is that the particle output distributions are aligned when the errors in the gap voltage lie within an acceptable range.

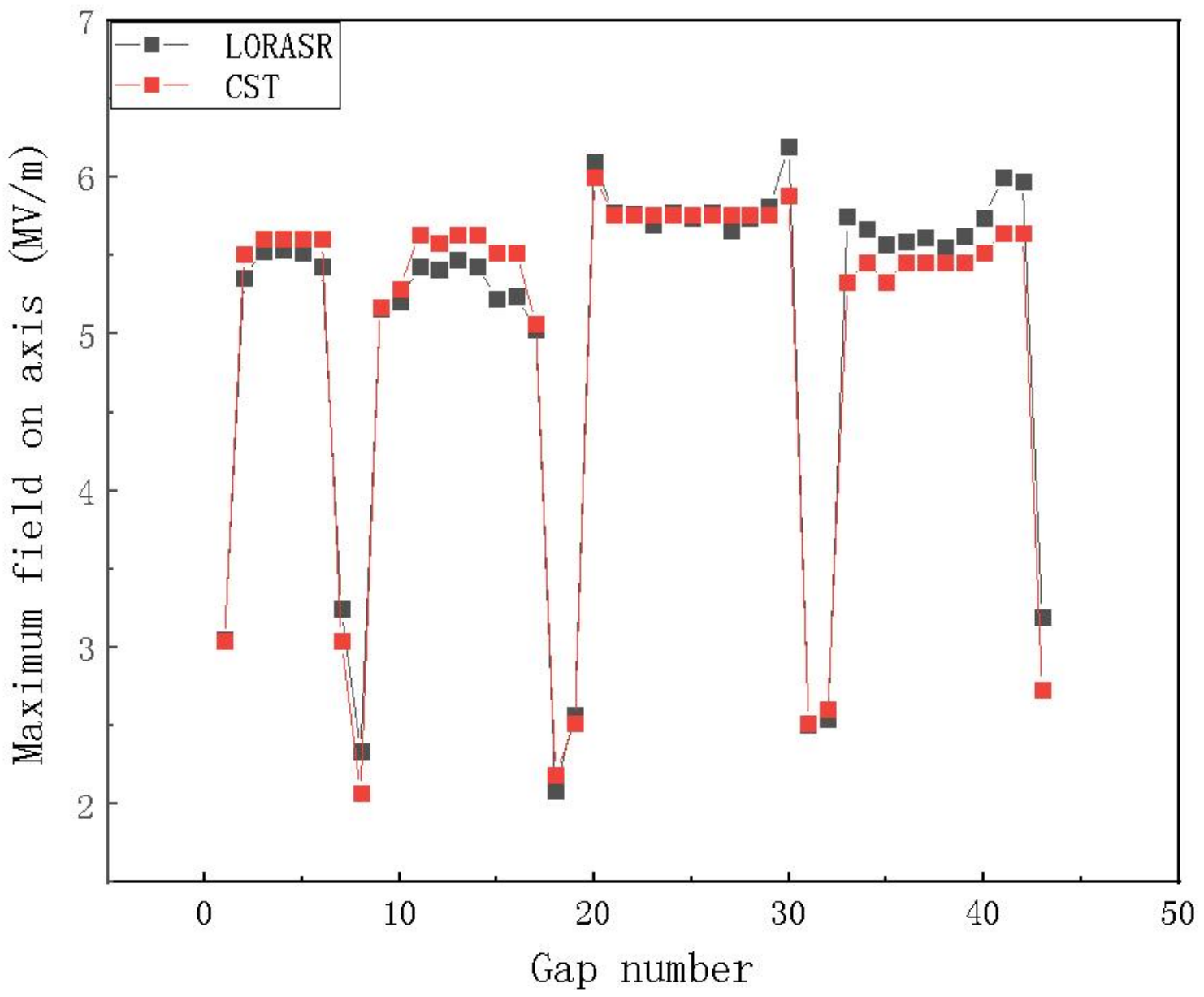

Figure 13. Maximum field on axis of dynamics and RF

The electric field distribution along the center axis of the dynamics design and the CST simulation are presented in Fig.13, the two results are basically consistent. For a relative effective voltage error

$$dU_{err}=(U_{Lo}-U_{cst})/U_{cst}$$

the distribution of the relative effective voltage error in each gap, shown in Fig.14, gives an error of less than 1%.

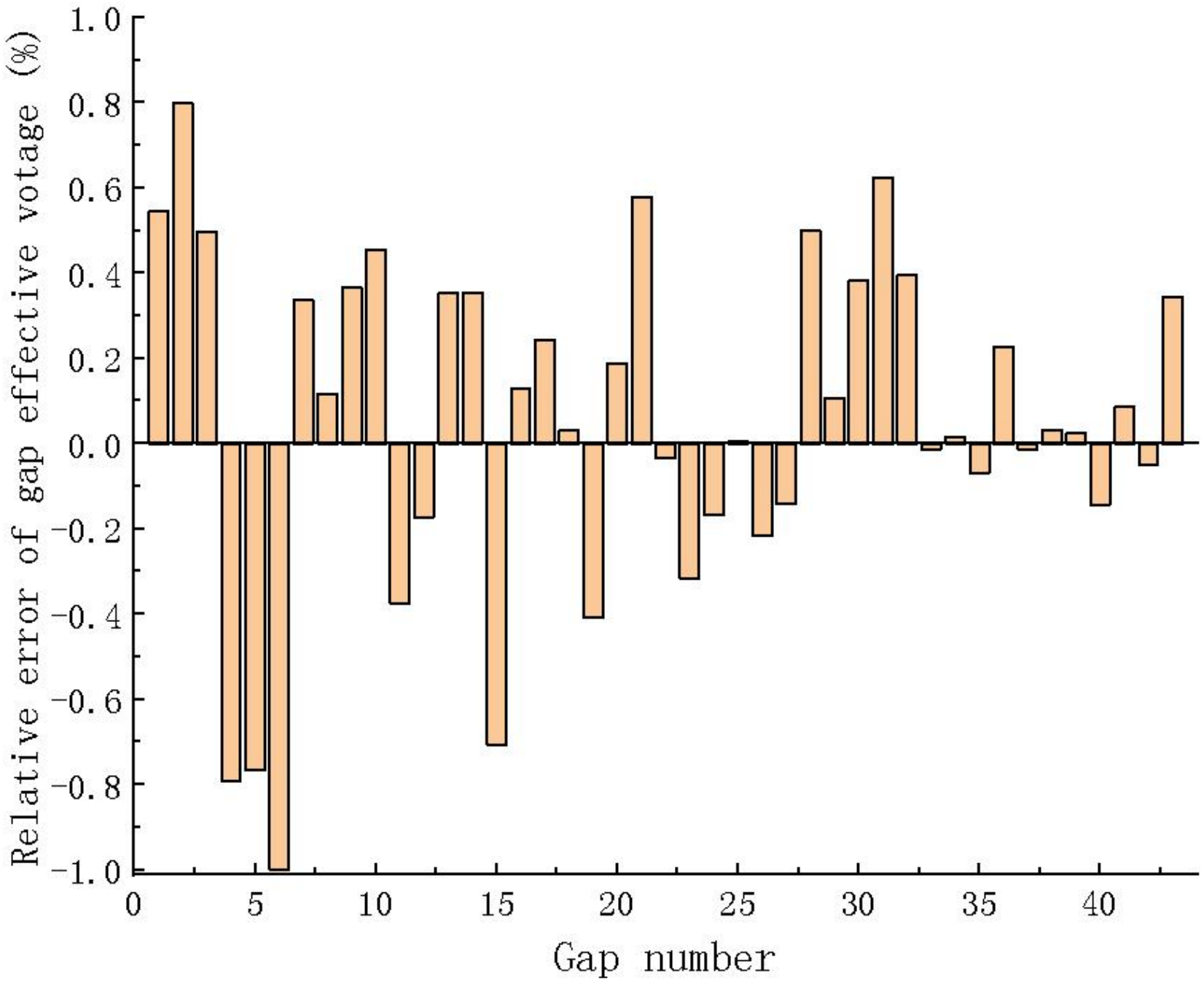

Figure 14 Relative error of effective voltage in each gap

The results obtained from the CST were further verified against the dynamics simulation, by importing the field distribution from CST into TraceWin [23] and comparing the results of the energy and emittance growth.

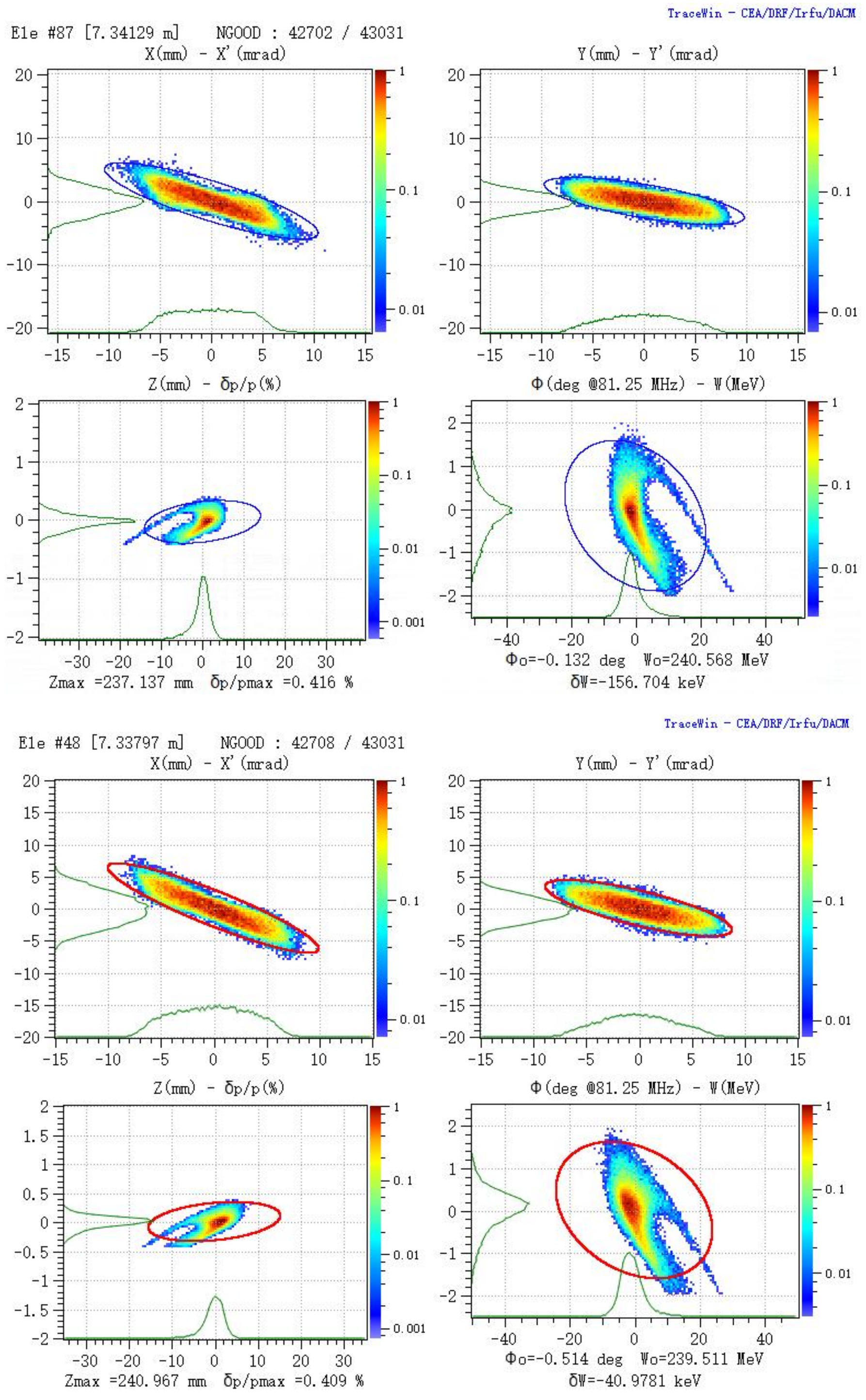

Figure 15. Beam output distributions for the dynamics (upper) and RF (lower) simulations.

The resulting beam output distributions, shown in Fig.15, match well in shape. The output parameters are summarized in Table 6. The output energy, transmission efficiency, and transverse normalized RMS emittance growth of the two simulations were all approximately equivalent. The longitudinal normalized RMS emittance in the RF simulation was slightly larger than in the dynamics simulation, but the results were consistent overall.

Table 6 Main output parameters of dynamics and RF

| Parameters | Dynamics | RF |
|---|---|---|
| Output energy [MeV/u] | 1.82 | 1.81 |
| Transmission/% | 99.2 | 99.2 |
| Output transverse normalized RMS emittance[$\pi$·mm·mrad] | 0.22/0.21 | 0.22/0.22 |
| Output longitudinal normalized RMS emittance[$\pi \cdot mm$·mrad] | 0.35 | 0.35 |

## 4.2 Error study

Error study is of course a critical component of particle accelerator design. Its purpose is to confirm whether the beams still accelerate effectively when the actual power deviates from the designed power, or when alignment errors are introduced into the cavities. It is more accurate to use the RF results obtained by CST for the real electromagnetic field distribution.

The comprehensive error analysis of the DTLs employed 100 beamlines with different error magnitudes. All error types of the corresponding maximum values are summarized in Table 7, including the input beam mismatch, voltage error of the cavity, alignment errors, and error of the QTs. For each beamline, the errors were uniformly distributed within the range of the maximum values. The input distribution utilized a macro-particle count of 43031, and the final transmission efficiency was 95.5%. The corresponding beam loss distribution, shown in Fig.16, indicates that most of the particle losses occurred in the QTs.

Table 7 Error range

| Error Type | Range |
|---|---|
| **Quadrupole Lenses** | |
| Singlet displacement(x/y) | 100 μm |
| Singlet rotation($\varphi_x/\varphi_y/\varphi_z$) | 1 deg |
| Gradient error | 1% |
| **Cavities** | |
| Cavity field | 1% |
| Cavity phase | 1 deg |
| Cavity displacement(x/y) | 100 μm |
| **Input Beam** | |
| Beam displacement(x/y) | 0.1 mm |
| Beam tilt($\varphi_x/\varphi_y$) | 2 mrad |
| Energy offset | 0.4 MeV |
| Twiss mismatch(x/y) | 5% |
| Beam current | 0.2 mA |

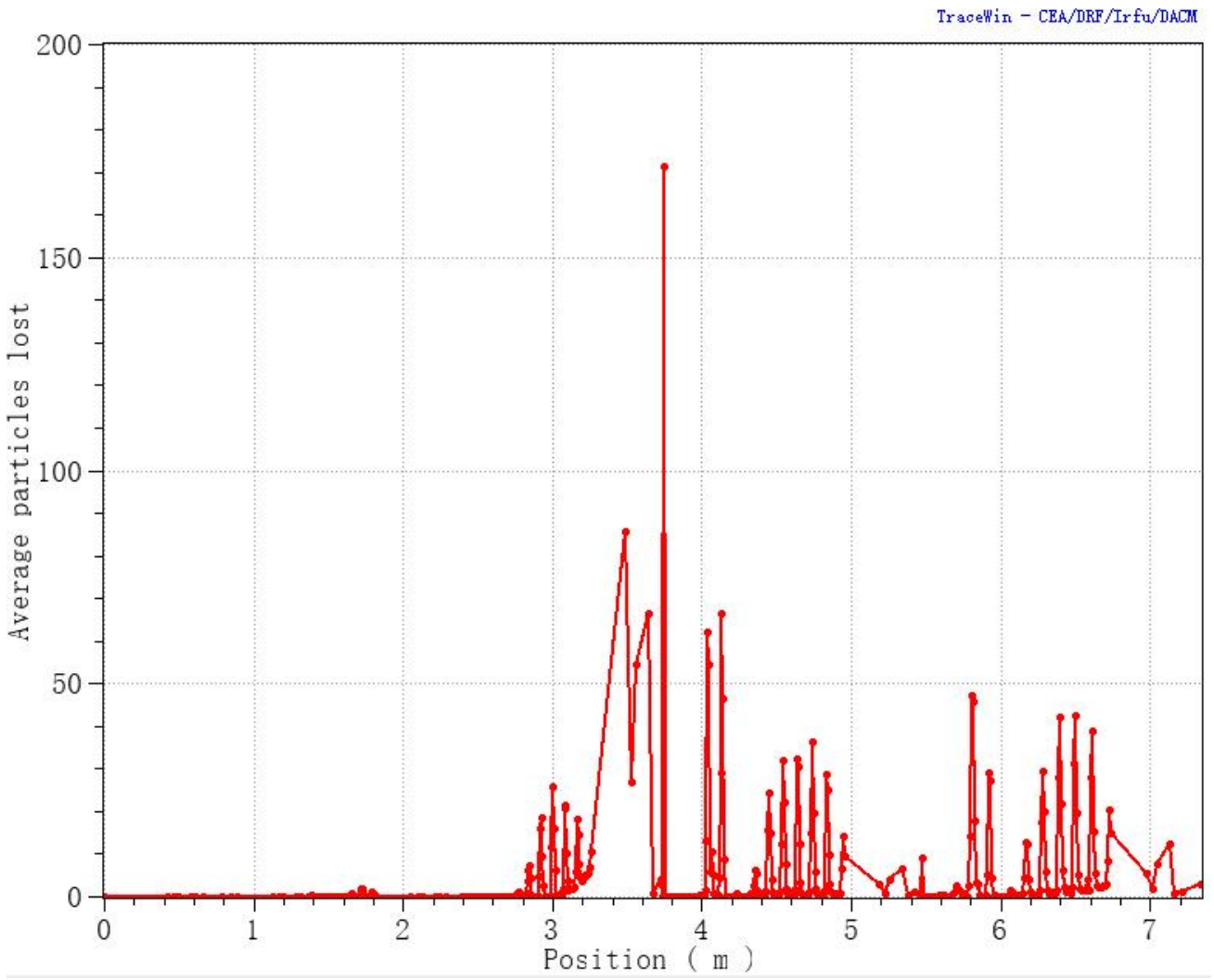

Figure 16 Average number of particles lost as a function of their position.

## Conclusion

In this paper, four 81.25 MHz CW CH-DTLs for the BISOL post-accelerator system were proposed and designed according to the new requirements. The dynamics design adopted the KONUS scheme utilizing the LORASR code, and indicated that the proposed DTLs are capable of accelerating $^{132}Sn^{22+}$ with 0.1 pmA from 0.5 MeV/u to 1.8 MeV/u over a length of 7 m, for an output longitudinal normalized RMS emittance of 0.35π·mm·mrad, and transmission efficiency higher than 95%. The employed CH structure in the electromagnetic design provides superior water-cooling characteristics for cavities lying within an optimal tuning range. The multiple-physics analysis found a maximum temperature rise of 7.7℃ in DTL3, maximum deformation and thermal stress in DTL1-Tank1, and a maximum frequency shift of -8.5kHz in DTL1-Tank2. All of the RF simulation results were within an acceptable range. The relative effective voltage error between dynamics and RF simulations was lower than 1%, and the results for both were determined to be consistent using TraceWin with respect to the beam output distribution and energy. Furthermore, the comprehensive error study implied that the design effectively accommodates non-ideal beams.